\documentclass[sn-mathphys]{sn-jnl}

\jyear{2021}%

\newcommand{\sty}[1]{\mbox{\boldmath $#1$}}

\newcommand{\fu}{\sty{ u}}
\newcommand{\ff}{\sty{ f}}
\newcommand{\fz}{\sty{ z}}
\newcommand{\fB}{\sty{ B}}
\newcommand{\fC}{\sty{ C}}

\newcommand{\fx}{\sty{ x}}
\newcommand{\fy}{\sty{ y}}
\newcommand{\fI}{\mathbb{\mathbf{I}}}
\newcommand{\feps}{\mbox{\boldmath $\varepsilon $}}
\newcommand{\fsig}{\mbox{\boldmath $\sigma$}}

\newcommand{\sumE}{\overset{m}{\underset{e=1}{\sum}}}
\newcommand{\argmin}{\mathop{\mathrm{arg\,min}}}
\DeclareFontFamily{U}{mathx}{\hyphenchar\font45}
\DeclareFontShape{U}{mathx}{m}{n}{
      <5> <6> <7> <8> <9> <10>
      <10.95> <12> <14.4> <17.28> <20.74> <24.88>
      mathx10
      }{}
\DeclareSymbolFont{mathx}{U}{mathx}{m}{n}
\DeclareMathSymbol{\bigtimes}{1}{mathx}{"91}

\algnewcommand\algorithmicbreak{\textbf{break}} 
\algnewcommand\Break{\algorithmicbreak{} }%
\algnewcommand\algorithmicdata{\textbf{Data:}}
\algnewcommand\Data{\item[\algorithmicdata{}]}%

\begin{document}

\title[Model-free Data-Driven simulation of inelastic materials]{Model-free Data-Driven simulation of inelastic materials using structured data sets, tangent space information and transition rules}

\author*[1]{\fnm{Kerem} \sur{Ciftci}}\email{Kerem.Ciftci@rub.de}

\author*[1]{\fnm{Klaus} \sur{Hackl}}\email{Klaus.Hackl@rub.de}

\affil[1]{\orgdiv{Institute of Mechanics of Materials}, \orgname{Ruhr-University Bochum}, \orgaddress{\city{Bochum}, \postcode{44801}, \country{Germany}}}


\abstract{Model-free data-driven computational mechanics replaces phenomenological constitutive functions by numerical simulations based on data sets of representative samples in stress-strain space. The distance of strain and stress pairs from the data set is minimized, subject to equilibrium and compatibility constraints. Although this method operates well for non-linear elastic problems, there are challenges dealing with history-dependent materials, since one and the same point in stress-strain space might correspond to different material behaviour. In recent literature, this issue has been treated by including local histories into the data set. However, there is still the necessity to include models for the evolution of specific internal variables. Thus, a mixed formulation  of classical and data-driven modeling is obtained. In the presented approach, the data set is augmented with directions in the tangent space of points in stress-strain space. Moreover, the data set is divided into subsets corresponding to different material behaviour. Based on this classification, transition rules map the modeling points to the various subsets. The approach will be applied to non-linear elasticity and elasto-plasticity with isotropic hardening.}




\maketitle

\section{Introduction}\label{sec:intro}
The simulation of boundary-value problems in solid mechanics typically combine two different types of equations; conservation and constitutive laws. The conservation laws are derived from universal principles containing an axiomatic character. Whereas the constitutive laws are formulated through modeling based on experimental observation. 
Material modeling aims to find these phenomenological models representing the data in the best way possible. Nevertheless, the process of modeling adds error and uncertainty to the solutions, especially in systems with high-dimensional complexity.\smallskip
\\ 
One approach to overcome this problem is the usage of machine learning techniques, especially artificial neural networks, to model material behaviour \cite{ghaboussi:1991,hkdh:1999,sha:2007}. The network is built directly from experimental data to recognize and learn the underlying nonlinear relations between strain and stresses without the construction of an explicit model. The performance of this approach is studied well for many kind of problems including plasticity \cite{zhang:2020}, high-performance material \cite{yeh:1998} and multiscale analysis \cite{unger:2009}. In relation, various neural network architectures have found applications in prediction \cite{huang:2002,yousif:2008,feng:2019}, modeling \cite{wang:2018,jones:2018,mozaffar:2019,vlassis:2020}, control and identification design \cite{koujelev:2010,greener:2018} areas of materials science. 
Despite their good reliability, neural networks rely on hidden layers. Therefore it is unclear on how much each independent variable is influencing the dependent variables, especially for higher-dimensional cases. \smallskip
\\ 
The model-free data-driven method by Kirchdoerfer and Ortiz \cite{kirchdoerfer:2016} incorporates experimental material data directly into numerical calculations of boundary-value problems. The method is based on a nearest neighbors approach. Particular in continuum mechanics, the optimization problem consists of calculating the closest point in the material data set consistent with the field equations of the problem i.e. compatibility and equilibrium. Therefore the data-driven method provides an alternative formulation of the classical initial-boundary-value problem completely bypassing the empirical material modeling step. For a variety of elasticity problems like (non-)linear material behaviour \cite{kirchdoerfer:2016,kirchdoerfer:2017,conti:2018,nguyen:2018,galetzka:2020}, dynamics \cite{kirchdoerfer:2018}, finite strain \cite{platzer:2021} and material data identification \cite{stainier:2019} the approach is elaborated and the associated convergence properties are well analyzed. However, problems arise when dealing with history-dependent data as present in inelastic materials, provides one uses nearest neighbor clustering only. Therefore, local histories are included into the data set in \cite{eggersmann:2019}. Nonetheless, it is still necessary to resort to additional models for the evolution of internal variables. Thus, a mixed formulation is obtained consisting of a combination of classical and data-driven modeling. In addition to that a variation of the scheme has been proposed considering multiscale modeling \cite{karapiperis:2021}. The framework emphasizes the parametrization of material history and the optimal sampling of the mechanical state space.\smallskip
\\
Recently, the data-driven scheme was extended by the tangent space, which improves the learning of the underlying data structure. One method is based on a manifold learning approach mapping the data into a lower-dimensional space to use locally linear embedding \cite{ibanez:2017,ibanez:2018}. A similar second order data-driven scheme, formulated by \cite{eggersmann:2021}, uses tensor voting \cite{mordohai:2010} to obtain point-wise tangent space. This enables the search for additional states close to the original data. Alternatively, this paper presents a the new approach by augmenting the tangent space directly into the distance-minimization data-driven formulation, which leads to a much more concise system of equations. Furthermore, the integration of the tangent space enables interpolation in regions of sparse data sampling, whilst ensuring the internal states to cohere with the data set. An additional step to deal with loading paths arising in inelasticity is to classify the underlying data structure into subsets corresponding to different material behaviour. Based on this, transition rules will be defined to map the internal states of the system to the various subsets. As a consequence, the extended data-driven paradigm evaluates the closest point in the transitioned material data subset consistent with the field equations of the problem and additionally closest to the local tangential direction. In the present study, we assume that all needed data is given. Furthermore, the question about data generation and accessibility from experiments remains open. This is a crucial topic that we are going to address in further research.
\smallskip
\\
To provide a general setting, Section~\ref{sec:datadriven} introduces the basic definitions and derivation of the classical distance-minimizing data-driven computing method. 
Section~\ref{sec:InelExt} presents the extension to inelasticity predicated on the extension of the data sets by tangent space information and the classification of the data into subsets corresponding to different material behaviour. Additionally transition rules are defined to map the modeling points to the various data subsets. Section~\ref{sec:NumExamples} demonstrates the performance of the suggested method via numerical examples employing non-linear elasticity and elasto-plasticity with isotropic hardening. At the end, Section~\ref{sec:conclusion} summarizes the results and gives recommendations concerning future research topics.

\section{Classical data-driven computing paradigm}\label{sec:datadriven} 
In the following the ordinary data-driven computational mechanics method will be summarized for the readers convenience based on the definitions and formulations in \cite{kirchdoerfer:2016,eggersmann:2019}.
Let $\Omega \subset \mathbb{R}^d$ with $d \in \mathbb{N}$ be a discretized system encountering displacements $\fu = \{\fu_i \in \mathbb{R}^{n_i}\}_{i=1}^n$ subjected to applied forces $\ff = \{\ff_i\in\mathbb{R}^{n_i} \}_{i=1}^n$, where  $n\in\mathbb{N}$ is the number of nodes and $n_i$ the dimension at node $i$.
The internal state is characterized by strain and stress pairs $\fz_e = (\feps_e,\fsig_e) \in  \mathbb{R}^{2d_e}$ with $d_e \in \mathbb{N}$ being the dimension of stress and strain at material point $e = 1,\ldots,m$, where $m\in \mathbb{N}$ is the number of material points.
The internal state of the system is subject to the compatibility and equilibrium condition
\begin{align}
&\feps_e = \fB_e \fu_e, \quad \forall e=1,\ldots,m, \label{eq:constraint1}\\ 
&\sumE  w_e \fB^T_e \fsig_e = \ff.  \label{eq:constraint2}
\end{align}
In this case $w_e$ is a positive weight and $\fB_e$ is a strain-displacement matrix.
Defining $\fz=\{(\feps_e,\fsig_e)\}_{e=1}^{m}$, the 
constraints \eqref{eq:constraint1} and \eqref{eq:constraint2} define a subspace %
\begin{equation}
\mathcal{C} :=\Big\{\fz\in\bigtimes_{e=1}^{m}\mathbb{R}^{2d_e}:\eqref{eq:constraint1}\; \text{and}\; \eqref{eq:constraint2}\Big\},
\end{equation}
denoted as constraint set with $\bigtimes$ being the Cartesian product.
Since the set is material-independent, the connectivity between $\feps_e$ and $\fsig_e$ is still missing. 
Instead of using a functional relationship, the information about the material is given by  means of a data set
\begin{equation}\label{eq:loc_dataset}
\mathcal{D} :=\Big\{\hat{\fz}\in\bigtimes_{e=1}^{m}\mathcal{D}_e\Big\}
\quad \text{with}  \quad
\mathcal{D}_e:= \{(\hat{\feps}_i, \hat{\fsig}_i) \in \mathbb{R}^{2d_e}\}_{i=1}^{n_e}, 
\end{equation}
where $n_e \in \mathbb{N}$ being the number of local data points; which classically consists of experimental measurements or data achieved from small scale simulations.
To define the data-driven problem the local space $\mathbb{R}^{2d_e}$ will be metricized by means of norms
\begin{equation}\label{eq:norm}
\|\fz_e\|_e := \frac{1}{2} E_e \|\feps_e\|^2_2 + \frac{1}{2} E_e^{-1} \|\fsig_e\|^2_2,
\end{equation}
with numerical scalar $E_e \in \mathbb{R}^+$, typically being of the type of an elastic stiffness, e.g., a representative Young's modulus. One might remark that this metric differs from the metric proposed in \cite{kirchdoerfer:2016}.
The corresponding local distance function
\begin{align}\label{eq:distance_e}
 d_e(\fz_e, \hat{\fz}_e ):= \|\fz_e - \hat{\fz}_e\|_e
\end{align}
with $\fz_e, \hat{\fz}_e \in \mathbb{R}^{2d_e}$, can be used to define a distance for $\fz, \hat{\fz} \in \bigtimes\limits_{e=1}^{m}\mathbb{R}^{2d_e}$ in the global space by
\begin{align}\label{eq:distance}
d(\fz, \hat{\fz}):=\sumE w_e d_e(\fz_e, \hat{\fz}_e ).
\end{align}
The distance-minimizing data-driven problem, introduced by \cite{kirchdoerfer:2016}, reads
\begin{equation}\label{eq:minmin}
\argmin_{\hat{\fz} \in \mathcal{D}} \argmin_{\fz\in \mathcal{C}}
 d(\fz, \hat{\fz}) =  \argmin_{\fz\in \mathcal{C}} \argmin_{\hat{\fz} \in \mathcal{D}}
 d(\fz, \hat{\fz}),
\end{equation}
i.e. the aim is to find the closest point consistent with the
kinematics and equilibrium laws to a material data set, or equivalently find the point in the data set that is closest to the constraint set.
The approach as well as the convergence and well-posedness have been studied on non-linear elastic material behaviour (cf. \cite{kirchdoerfer:2016,conti:2018}). In the following the data-driven paradigm will be extended by the tangent space.   

 \section{Extension by tangent space}\label{sec:InelExt}
In the following, we will suggest an extension of the data-driven paradigm by including tangent space information in order to deal with inelastic materials. This is a non-trivial task, since the same point in stress-strain space might correspond to different material behavior. 
Whereas it is proposed in \cite{eggersmann:2019} to include local histories into the data set, we will extend the data set by the tangent space information. 
For this purpose, let us introduce the extended data set 
\begin{align}
\mathcal{D}^{\text{ext}} = \bigtimes_{e=1}^m \mathcal{D}_e^{\text{ext}} \quad
\text{with} \quad \mathcal{D}_e^{\text{ext}}:=\{(\hat{\fz}_i, \fC_i) \,\vert\,  \hat{\fz}_i\in \mathcal{D}_e, \fC_i \in \mathbb{R}^{d_e \times d_e}\}_{i=1}^{n_e},
\end{align}
where $\fC_i$ represents the total stiffness matrix at $(\hat{\feps}_i, \hat{\fsig}_i)$, including potential inelastic effects. Thus, the actual independent data is given by $((\hat{\feps}_i, \hat{\fsig}_i),\fC_i)$, i.e. strain, stress and stiffness matrix. We are fully aware, that measuring $\fC_i$ experimentally might be a formidable task. However, it might very well be possible combining information on nearby strain and stress pairs and employing material symmetry. We plan to elaborate on this in a subsequent paper. For now, we will simply assume $\fC_i$ to be available.
\\
The tangent space extension allows to operate on the underlying structure of the phase space of strain and stress pairs. In the following, we will introduce a way to incorporate the tangent space directly into the data-driven computing paradigm. 
\subsection{Data-driven formulation}
\begin{figure}[h]
	\includegraphics[scale=0.75]{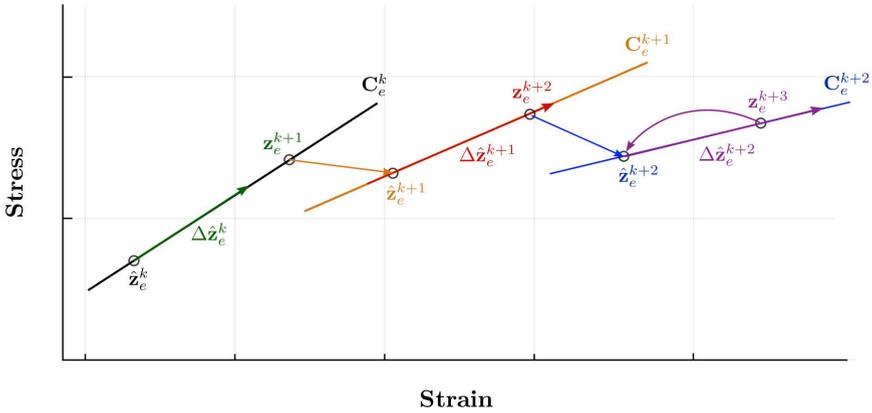}
	\caption{Visualization of data-driven method extended by tangent space. Modeling points $\fz_e^{k+1}$ minimize distance to the tangent space associated with data points $\hat{\fz}_e^k$, respecting compatibility and equilibrium constraints. Data points $\hat{\fz}_{e}^{k+1}$ minimize distance to modeling points $\fz_e^{k+1}$. Iterations are repeated until the local data assignments remain unchanged or the global distance is less than a certain tolerance.}
	\label{fig:dditeration}
\end{figure}

Recalling the distance-minimization problem \eqref{eq:minmin}, we start by evaluating the data point $(\hat{\fz}, \fC) = \{(\hat{\fz}_e, \fC_e)\}_{e=1}^m$ in the extended data set closest to the constraint set, i.e.
\begin{align}
 \argmin_{(\hat{\fz}, \fC) \in \mathcal{D}^{\text{ext}}} d(\mathcal{C},\hat{\fz}).
\end{align}
Then, each local optimal data point and its corresponding tangent can then be used to define a map
$\fy_e:\mathbb{R}^{d_e}\to\mathbb{R}^{d_e}$ with 
\begin{align}
 \fy_e(\fx_e)  =  \hat{\fsig}_e + \fC_e ( \fx_e -\hat{\feps}_e)  \qquad \forall e=1,\ldots,m,
\end{align}
parametrizing the tangent space as subset of the phase space. Thus, the data sets on which the data-driven paradigm operates can be written as
\begin{align}
\mathcal{D}^{\Delta} = \bigtimes_{e=1}^m \mathcal{D}_e^{\Delta} \quad
\text{with}  \quad \mathcal{D}_e^{\Delta}:= \{(\fx_e, \fy_e(\fx_e)) \,\vert\, \fx_e \in \mathbb{R}^{d_e}\}.
\end{align}
This definition allows to incorporate the local tangent spaces directly into the distance-minimization formulation, i.e.
\begin{align}\label{eq:minC}
\argmin_{\fz \in \mathcal{C}} d(\fz, \mathcal{D}^\Delta),
\end{align}
using the underlying data structure. For this purpose, the remaining step is the determination of the  material state  $\fz = \{(\feps_e,\fsig_e)\}_{e=1}^{m} \in \mathcal{C}$ closest to the data sets $\mathcal{D}^{\Delta}$. For given optimal data sets $\Delta\hat{\fz}_e = (\fx_e, \fy_e(\fx_e))\in \mathcal{D}_e^\Delta$, e.g. from a previous iteration, the minimization problem~\eqref{eq:minC} can then be written as
\begin{equation}\label{eq:minprob}
\begin{aligned}
 \textrm{Minimize} \quad &  \sumE w_e d_e(\fz_e, \Delta\hat{\fz}_e) 
\\
\textrm{s.t.} \quad & \feps_e =\fB_e \fu_e \quad \text{and } \quad
\sumE  w_e \fB^T_e \fsig_e = \ff.
\end{aligned}
\end{equation}
Taking the Cartesian product structure of $\fz \in \bigtimes\limits_{e=1}^{m}\mathbb{R}^{2d_e}$ into account, it is reasonable to interchange the summation and the minimization. Therefore it is enough to calculate 
\begin{align}
\min\limits_{\fz_e  \in \mathbb{R}^{2d_e}} d_e(\fz_e, \Delta \hat{\fz}_e) = \min\limits_{\fz_e  \in \mathbb{R}^{2d_e}} \|\fz_e - \Delta \hat{\fz}_e\|_e  \qquad \forall e=1,\ldots,m,
\end{align}
where we used the definition of the distance function.
One can notice that the minimizing state can be found by evaluating the material state $\fz_e =  \Delta \hat{\fz}_e$.
Enforcing the compatibility constraint by expressing the material strains in terms of displacements $\fz_e = (\fB_e \fu_e,\fsig_e)$ it follows
\begin{align}
\feps_e &= \fB_e \fu_e = \fx_e  && \forall e=1,\ldots,m, \label{eq:eps_e}\\
\fsig_e &=  \fy_e(\fx_e) = \hat{\fsig}_e + \fC_e ( \fx_e -\hat{\feps}_e) && \forall e=1,\ldots,m. \label{eq:sig_e}
\end{align}
Substitution of \eqref{eq:eps_e} into \eqref{eq:sig_e} leads to 
\begin{align}
\fsig_e =  \hat{\fsig}_e + \fC_e (\fB_e \fu_e - \hat{\feps}_e)  \qquad  & \quad && \forall e=1,\ldots,m.
\end{align}
Now using the equilibrium constraint we have
\begin{align}
\sumE  w_e \fB^T_e \fsig_e = \sumE  w_e \fB^T_e(  \hat{\fsig}_e + \fC_e (\fB_e \fu_e - \hat{\feps}_e)) = \ff.
\end{align}
Finally reordering leads to a standard linear problem given by
\begin{equation}\label{eq:lin_eq_sys}
\left(\sumE  w_e \fB^T_e \fC_e \fB_e \right) \fu = \ff - \sumE  w_e \fB^T_e (\hat{\fsig}_e  - \fC_e \hat{\feps}_e ).
\end{equation}
Solving the equation system for $\fu$ and make use of \eqref{eq:eps_e} and \eqref{eq:sig_e}, we can calculate the closest local material states to the local data sets ensuring the compatibility and equilibrium condition.
\\
Since we assumed given optimal data points, it remains to determine the stress, strain and tangent space pairs $(\hat{\fz}_e, \fC_e)$ in the local data sets $\mathcal{D}^{\text{ext}}_e$ that result in the closest possible satisfaction of compatibility and equilibrium. 
The determination of the optimal points can be done iteratively. For given data points $\{(\hat{\fz}_e^k, \fC_e^k)\}_{e=1}^m$ at iteration $k$ the modeling points $\{\fz_e^{k+1}\}_{e=1}^m$ are calculated using the data-driven scheme. Next, we calculate the closest local data points in the extended set to the latest modeling points. The iterations are performed until the data assignment remain unchanged or the global distance $d(\fz, \hat{\fz})$ is lower than a predefined tolerance, we reached convergence. A visualization of a single algorithmic loading step is given in Fig.~\ref{fig:dditeration} and the detailed extended data-driven scheme is summarized in Algorithm \ref{alg:ddsolver_nonlin}.
Due to the usage of the tangent-space structure, only a few or even just one iteration are required.
This constitutes a considerable increase of efficiency in comparison with the traditional data-driven algorithm introduced in \cite{kirchdoerfer:2016}. We realize that the issue about the accessibility of data and its corresponding tangent space is crucial. As mentioned before, we are assuming that all required data is given. The question about data generation and availability will be addressed in further research. 
\begin{algorithm}[H]
\caption{Extended data-driven solver}
\label{alg:ddsolver_nonlin}
\begin{algorithmic}
\Require matrices $\{\fB_e\}_{e=1}^m$, weights $\{w_e\}_{e=1}^m$, load $\ff$, tolerance $\mathrm{tol}$
\vspace{0.3cm}
\Data data points $\{(\hat{\fz}_e, \fC_e)\}_{e=1}^m$, data sets $\{\mathcal{D}_e^{\text{ext}}\}_{e=1}^m$, 
\vspace{0.3cm}
\Function{DDsolver}{$\{\mathcal{D}_e^{\text{ext}}\}_{e=1}^m, \{(\hat{\fz}_e, \fC_e)\}_{e=1}^m, \ff$}
\State Set iteration $k = 0$
\State $\{(\fz_e^k, \fC^k)\}_{e=1}^m\leftarrow\{(\hat{\fz}_e, \fC_e)\}_{e=1}^m$
\While {true}
\State Solve equation system: 
\begin{align*}
\left(\sumE  w_e \fB^T_e \fC_e^k \fB_e \right) \fu^{k+1} = \ff^{k+1} - \sumE  w_e \fB^T_e (\hat{\fsig}_e^k - \fC_e^k \hat{\feps}_e^k)
\end{align*}
\For{$e = 1 \to m$}
\State $\feps_e^{k+1} = \fB_e \fu^{k+1},$
\State $\fsig_e^{k+1} = \hat{\fsig}_e^k + \fC_e^k(\feps_e^{k+1} - \hat{\feps}_e^k)$
\EndFor
\For{$e=1 \to m$}
\State $\min\{ d_e(\fz_e^{k+1},\hat{\fz}_{e}^{k+1})\,\vert\, ({\hat{\fz}_{e}, \fC_e) \in \mathcal{D}}_e^{\text{ext}}\}$ 
\EndFor
\If{$d(\fz^{k+1}, \hat{\fz}^{k+1}) \leq \mathrm{tol}$}
\State $\{\fz_e\}_{e=1}^m \leftarrow \{\fz_e^{k+1}\}_{e=1}^m$
\State \Break
\Else
\State $k \leftarrow k + 1$
\EndIf
\EndWhile
\State \Return $ \{\fz_e\}_{e=1}^m$
\EndFunction
  \end{algorithmic}
\end{algorithm}

\subsection{Transition rules}
To simulate inelastic material behaviour, the main task is to capture history dependence. 
This is achieved by associating different tangent spaces to data points with different history. 
Assuming an underlying data structure, as proposed in \cite{eggersmann:2021}, the local material data sets $\mathcal{D}_{e}^{\text{ext}}$ are classified into subsets corresponding to different material behaviour, e.g. elastic and inelastic:
\begin{align}
\mathcal{D}_e^\text{ext} =  \dot{\bigcup\limits_p}\, \mathcal{D}_e^{\text{ext},\,p} \quad \text{with } p = \{\text{elastic},\,\text{inelastic}\}. 
\end{align}
Thus, data points with close or even the same strain and stress values may possess vastly different tangent spaces; in the elastic case essentially determined by the elastic stiffness and in the plastic case by the hardening modulus. 
It should be emphasized that it is easily possible to distinguish experimentally between elastic and plastic material behaviour.
Based on the classification, transition rules map the modeling points to the various subsets. 
\\
In the following, a transition mapping is derived for the case of elasto-plasticity with isotropic hardening. The kinematics of elasto-plasticity is governed by a yield condition of the form
\begin{equation}
\label{yield}
\sigma_\mathrm{com}(\fsig) \leq \sigma_\mathrm{y},
\end{equation}
where $\sigma_\mathrm{com}(\fsig)$ is a comparison stress dependent on the current stress state, e.g. $\sigma_\mathrm{com}(\fsig)=\sqrt{3/2} \, \| \mathrm{dev} \fsig\|$ in the case of von Mises ($J_2$) plasticity, and $\sigma_\mathrm{y}$ denotes the yield stress, a material property depending on the loading history in the case of isotropic hardening. 
For $\sigma_\mathrm{com}(\fsig) < \sigma_\mathrm{y}$, we have elastic behaviour, for $\sigma_\mathrm{com}(\fsig) = \sigma_\mathrm{y}$ plastic behaviour.
\\ \\
Given values of modeling points $\{\fz_e\}_{e=1}^m$ using the data-driven algorithm \ref{alg:ddsolver_nonlin}, the transition mapping for material state $e=1,\ldots,m$ at time step $t+1$ can be formulated as:
\begin{enumerate}
\item \label{itm:dataset} assign local data set $\tilde{\mathcal{D}}_e^\text{ext}$ by
\begin{align}
\tilde{\mathcal{D}}_e^\text{ext} = \mathcal{D}_e^{\text{ext},\, p} \quad \text{with } 
p \equiv
\begin{cases}
 \text{elastic}, &\text{if }
\sigma_\mathrm{com} ({\fsig_e}) < \sigma_{\mathrm{y},e} \\ 
\text{inelastic}, &\text{otherwise}.
\end{cases}
\end{align}
\item\label{itm:yield} if $p \equiv \text{inelastic}$, set new yield stress at $\sigma_{\mathrm{y},e} := \sigma_\mathrm{com} ({\fsig_e})$;
\item \label{itm:nneighbor} find closest data point $\{(\hat{\fz}_e, \fC_e)\}_{e=1}^m$ in data set $\tilde{\mathcal{D}}_e^\Delta$ to modeling point $\fz_e$ by
\begin{align}
\min\{ d_e(\fz_e,\hat{\fz}_{e})\,\vert\, (\hat{\fz}_e, \fC_e)\in \tilde{\mathcal{D}}_e^\text{ext} \}.
\end{align}
\end{enumerate}
While step \ref{itm:dataset} maps the modeling points to the corresponding data sets, steps \ref{itm:yield} and \ref{itm:nneighbor} define a new yield limit and find the closest data point inside these sets for the next loading increment. 
These formulations give rise to corresponding representational scheme in the context of data-driven inelasticity, which are summarized in Algorithm \ref{alg:ddsolver_inelastic}.
\begin{algorithm}[H]
\caption{Data-driven transition rules for inelasticity at time step $t+1$}
\label{alg:ddsolver_inelastic}
\begin{algorithmic}
\Require load $\ff$, yield stresses $\{\sigma_{\mathrm{y},e}\}_{e=1}^m$
\vspace{0.15cm}
\Data data points $\{(\hat{\fz}_e, \fC_e)\}_{e=1}^m$, data subsets $\{(\mathcal{D}_e^{\text{ext},\,\text{elastic}}, \mathcal{D}_e^{\text{ext},\,\text{inelastic}})\}_{e=1}^m $
\vspace{0.3cm}
\State $\{\fz_e\}_{e=1}^m$ = \Call{DDsolver}{$\{ \tilde{\mathcal{D}}_e^\text{ext}\}_{e=1}^m, \{(\hat{\fz}_e, \fC_e)\}_{e=1}^m, \ff$}
\vspace{0.3cm}
\For{$e=1 \to m$}
\If{$\sigma_\mathrm{com} ({\fsig_e}) < \sigma_{\mathrm{y},e}$} 
\State $\tilde{\mathcal{D}}_e^\text{ext} \equiv \mathcal{D}_e^{\text{ext},\,\text{elastic}}$
\Else
\State $\tilde{\mathcal{D}}_e^\text{ext} \equiv \mathcal{D}_e^{\text{ext},\,\text{inelastic}}$
\State $\sigma_{\mathrm{y},e} = \sigma_\mathrm{com} ({\fsig_e})$ 
\EndIf
\State $\min\{ d_e(\fz_e,\hat{\fz}_{e})\,\vert\, (\hat{\fz}_e, \fC_e) \in \tilde{\mathcal{D}}_e^\text{ext} \}$ 
\EndFor
\State $t+1 \leftarrow t+2$
\end{algorithmic}
\end{algorithm}
\section{Numerical examples}\label{sec:NumExamples}
In this section the performance of the presented data-driven solver extended by the tangential space information will be illustrated in two typical benchmark examples considering the stress analysis of  non-linear elastic material and an elasto-plastic von Mises material with isotropic hardening. In this scope, we discuss the accuracy and convergence. 
\\  
The error between a data-driven solution $\fz^k$ and its corresponding reference solution $\fz^{k, \text{ref}}$ shall be calculated by means of the root-mean-square deviation of strain and stress defined by
\begin{align}
\text{RMSD}(\fz)^2 &= \frac{\sum_{k=0}^T \text{Error}(\fz^k )^2}{T},
\end{align}
where $T \in \mathbb{N}$ is the number of total loading steps, $\fz^k_e = (\feps_e^k, \fsig_e^k )$ the local data-driven states and  $\fz^{k, \text{ref}}_e = (\feps_e^{k, ref} ,\fsig_e^{k, \text{ref}})$ the local reference states at step $k \leq T$. The error is given by 
\begin{align}
\text{Error}(\fz^k )^2 &=  \frac{\sum_{e=1}^m w_e \|\fz_e^k - \fz^{k, \text{ref}}_e\|^2}{ \sum_{e=1}^m w_e \|\fz^{k, \text{ref}}_e\|^2},
\end{align}
with $\|\cdot\|$ given by definition \eqref{eq:norm}. 
\subsection{Non-linear elastic cylindrical tube under internal pressure}
\begin{figure}[H]
\includegraphics[scale=0.19]{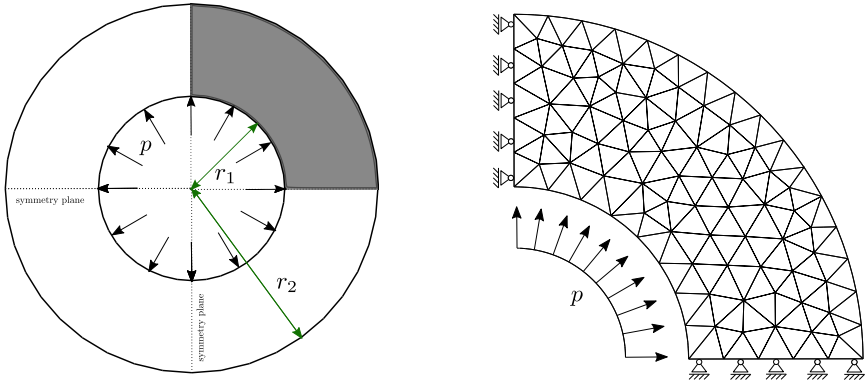}
\caption{Discretization and boundary conditions for a rectangular plate with a circular hole under loading.}
\label{fig:cylinder_geometry}
\end{figure}

The first example is the classical benchmark problem  considering a non-linear elastic cylindrical tube under internal pressure $p$. Fig.~\ref{fig:cylinder_geometry} illustrates the geometry and the boundary conditions for this particular problem
The tube has thickness $t = r_2 - r_2$ with inner and outer radii $r_1 = 1\mathrm{m}$ and $r_2 = 2\mathrm{m}$.
Two symmetry planes can be identified and therefore the solution domain need only cover a quarter of the geometry, shown by the shaded area. 
The domain is discretized by quadratic triangles. 
Modeling the tube as a two dimensional plane-strain problem, the corresponding material parameters of the reference solid used for the reference solution and data sets are Young's modulus $E = 70\cdot 10^3\, \text{Pa}$, Poisson's ratio $\nu = 0.3$ and elasticity tensor
\begin{align}\label{eq:elas_matrix}
\mathbb{C} = \lambda \fI \otimes \fI + 2\mu \mathbf{\mathbb{I}},
\end{align}
where $\fI$ is the second-rank identity tensor, $\mathbf{\mathbb{I}}$ is the symmetric part of the fourth-rank identity tensor and $\lambda = \frac{E\nu}{(1+\nu)(1-2\nu)}$ and $\mu = \frac{E}{2(1+\nu)}$ are the Lamé constants.
The response is computed using a non-linear relation
\begin{align}\label{eq:nl_model}
\fsig(\feps) = \lambda f(\mathrm{tr}(\feps))\fI + \mu \feps + \mathbb{C} : \feps
\end{align}
with $f(x):=c_1\arctan(c_2 x)$, parameters $c_1 = 3.0 \cdot 10^{-2}$, $c_2 = 1.0\cdot 10^2$.

For the data-driven computation two different types of data distributions are investigated. 
The first data set is created by a zero-mean normal distribution with a standard deviation of $0.01$ and the second data set is created by a uniform distribution within $[-0.02,0.02]$ for strains in each direction. 
The corresponding local tangents $\mathcal{C}$ are calculated analytically using equation \eqref{eq:nl_model} and endued with some noise. 
Finally, the simulation of problem in Fig.~\ref{fig:cylinder_geometry} is performed by applying a pressure $p(t) = \dfrac{5\cdot 10^4}{\sqrt{3}} \log\left(\dfrac{r_2}{r_1}\right)\cdot t$ progressively increased with $100$ incremental steps using a constant normalized time step of $\Delta t = 1$. 
Due to the random nature of the data distribution, each simulation returns a different error. 
To cover a wide spectrum of the errors produced, we run $100$ simulations corresponding to independent realizations of both, normal and uniform distribution.
\begin{figure}[H]
    \subfloat[]{{\includegraphics[scale=0.32]{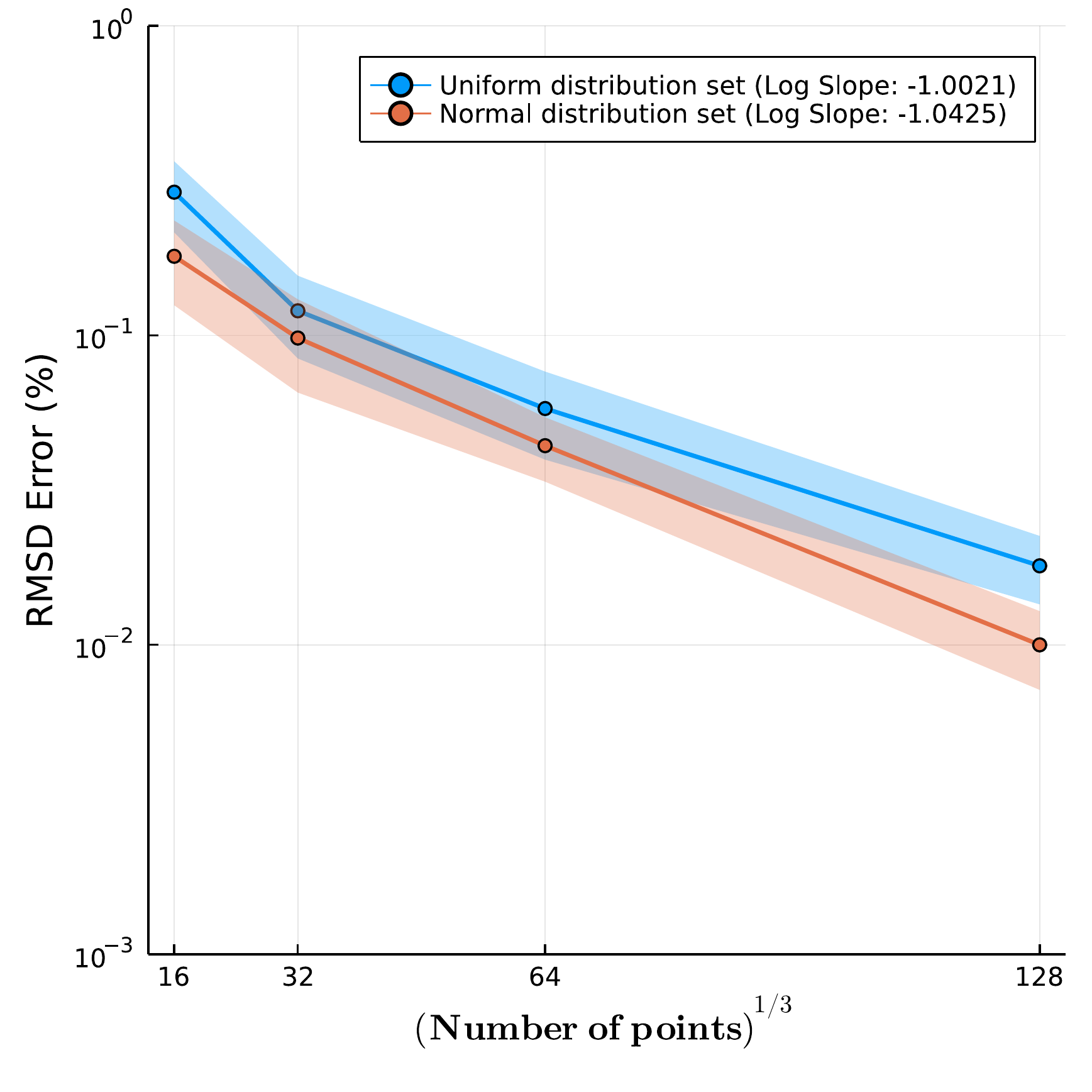} }\label{fig:conv_nl_log}}%
    \qquad
    \subfloat[]{{\includegraphics[scale=0.32]{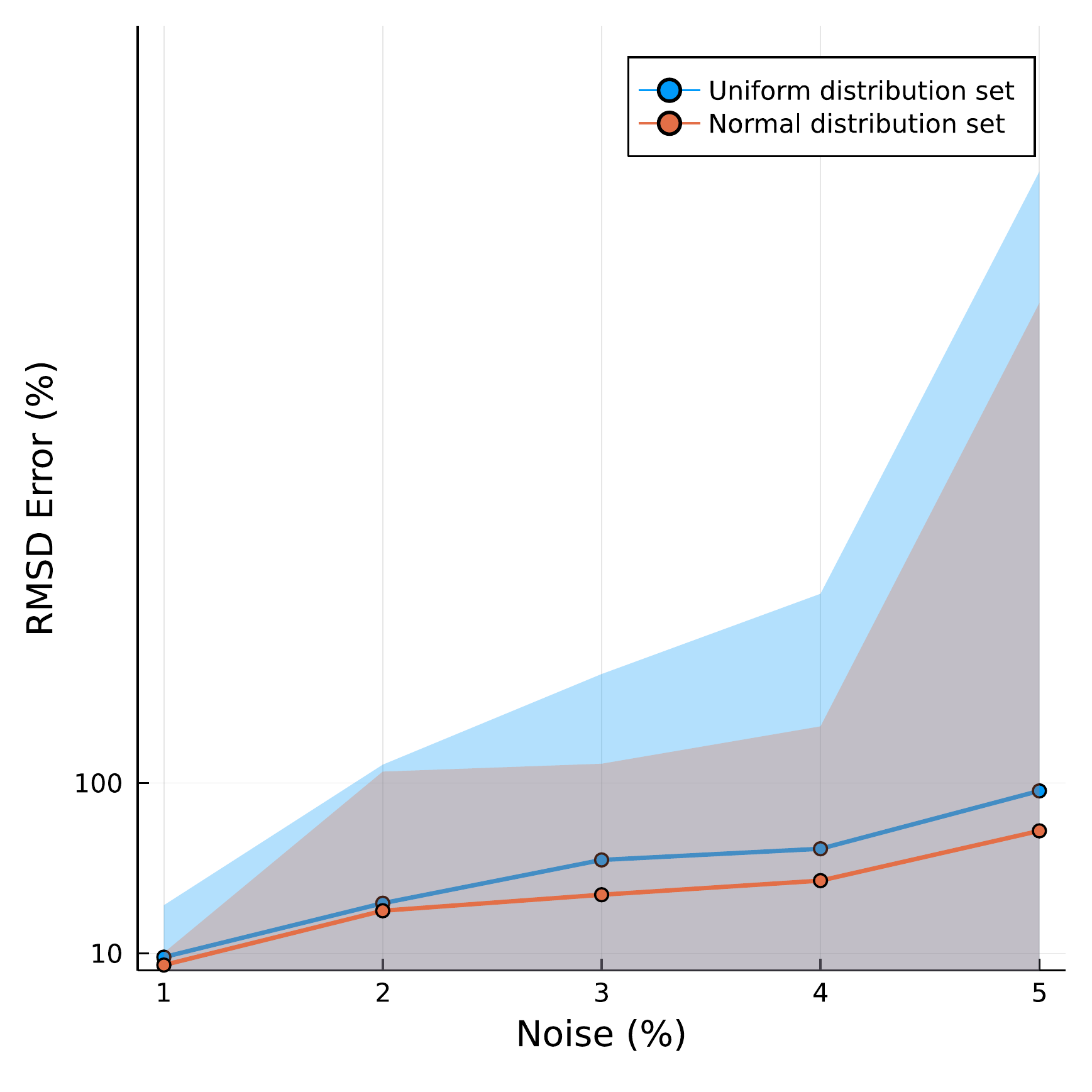} }%
    \label{fig:conv_nl_dev}}
    \caption{RMSD Error of data-driven solver for normal and uniform distributed data points. (a) Convergence with respect to data size. (b) Dependency of the error on applied noising for a data set of size $16^3$ with normal and uniform distribution. The shaded areas show the spread of the error arising from the different data set realizations.}%
    \label{fig:nl_example}%
\end{figure}
The error plot in Fig.~\ref{fig:conv_nl_log} shows a linear rate of convergence, which corresponds to the data-driven convergence analysis of elastic problems in \cite{kirchdoerfer:2016}. 
Figure~\ref{fig:conv_nl_dev} shows the dependence of the error from noising ranging from $1\%$ to $10\%$ of the maximum values of strains and stresses applied to the various data sets.
The shaded areas show the spread of the error arising from the different data set realizations used in the independent simulation runs. 
Note, that apparently data sets with normal distribution give better performance.
\begin{figure}[H]
    \subfloat[]{{\includegraphics[scale=0.26]{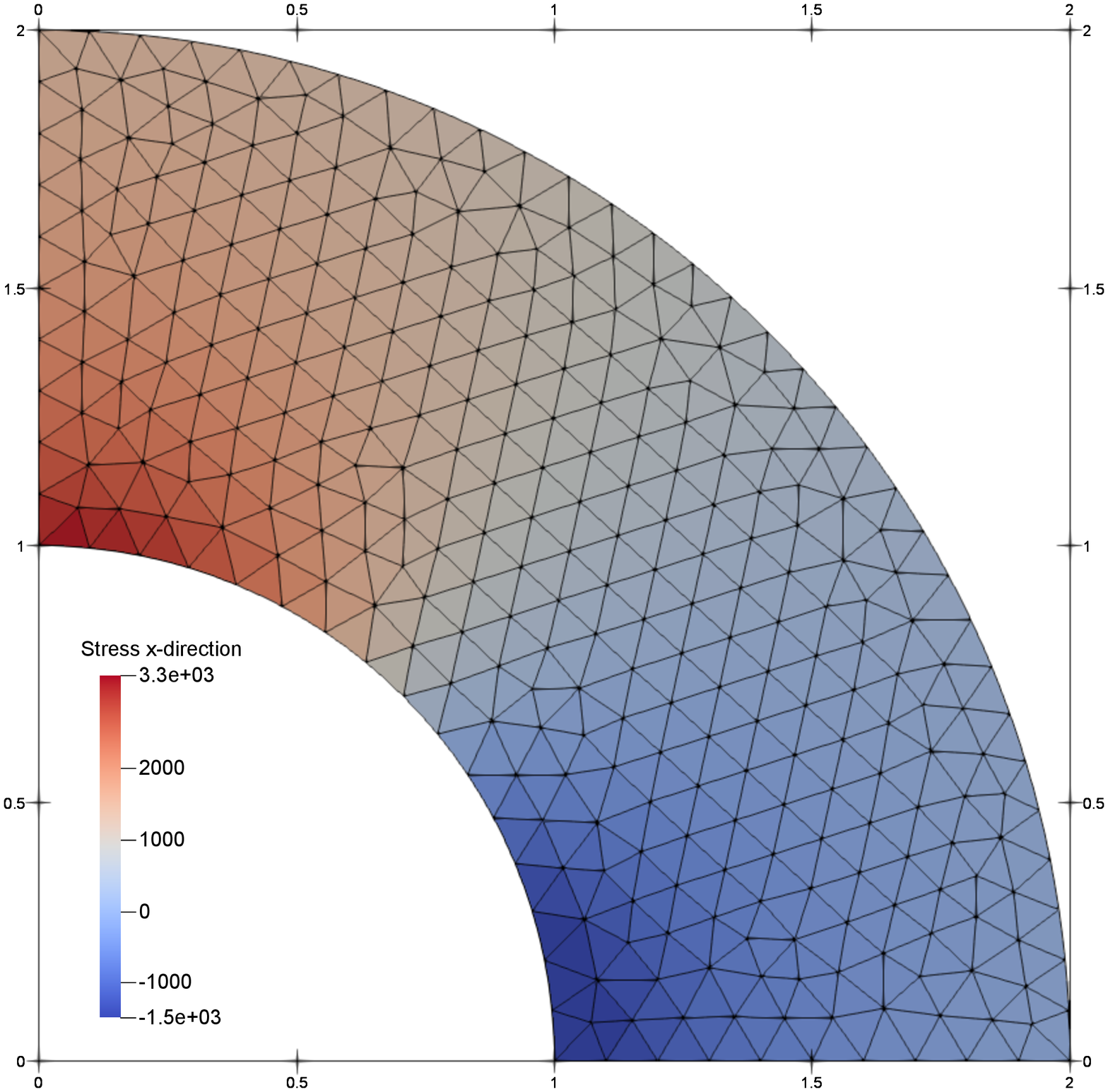} }\label{fig:nl_sig_xx}}%
    \qquad
    \subfloat[]{{\includegraphics[scale=0.26]{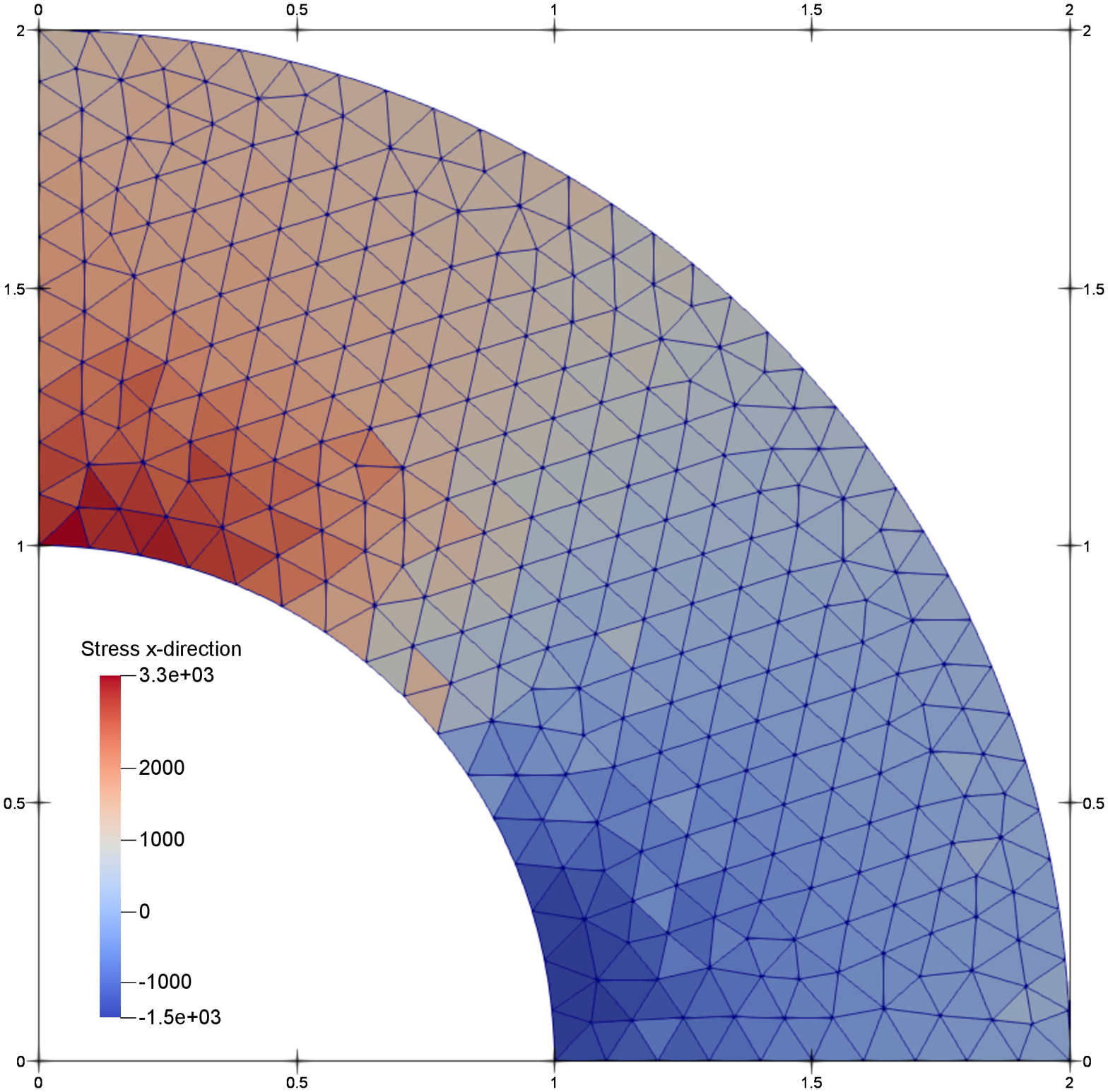} }%
    \label{fig:nl_ddsig_xx}} \\
    \subfloat[]{{\includegraphics[scale=0.26]{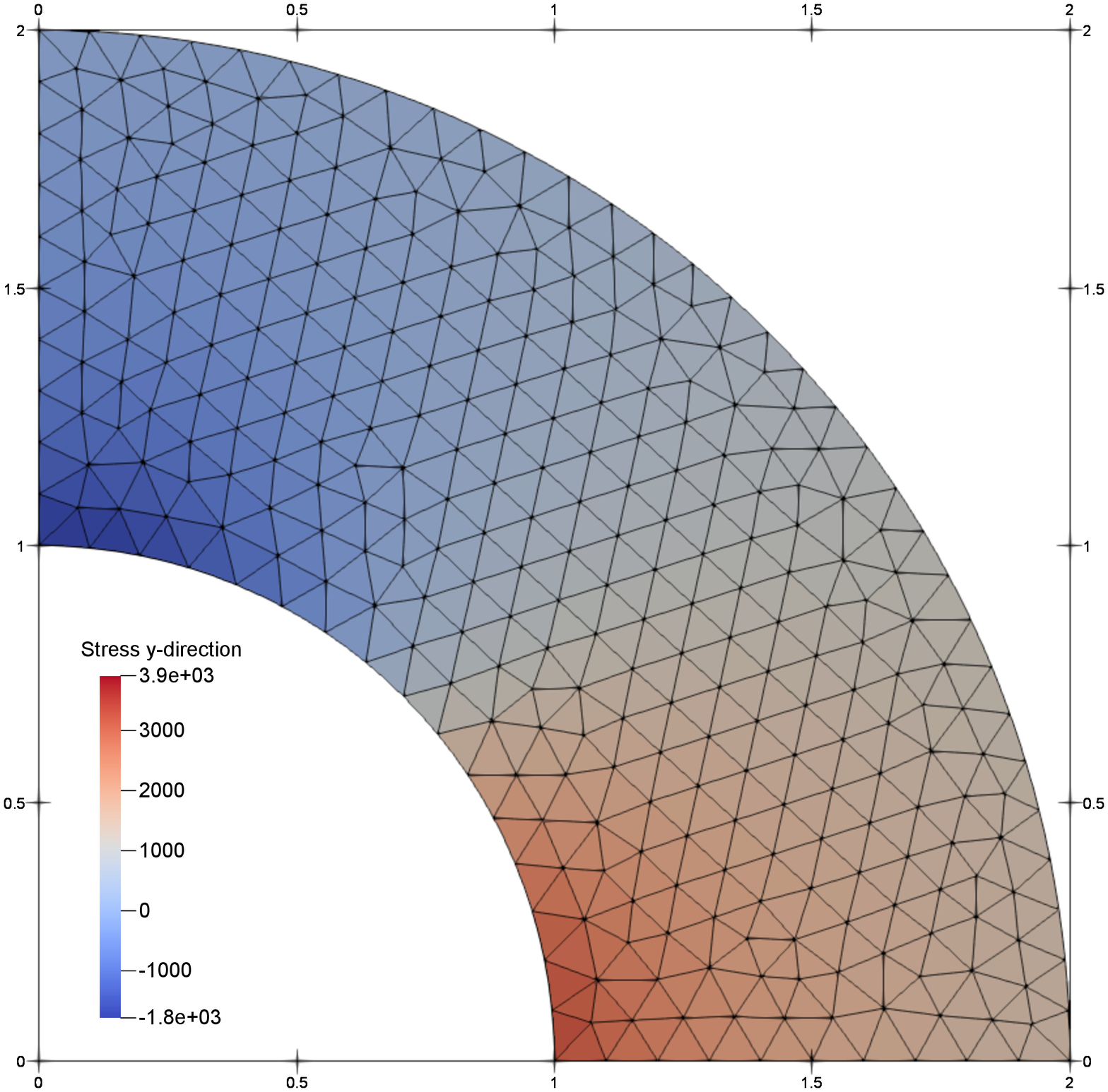} }\label{fig:nl_sig_yy}}%
    \qquad
    \subfloat[]{{\includegraphics[scale=0.26]{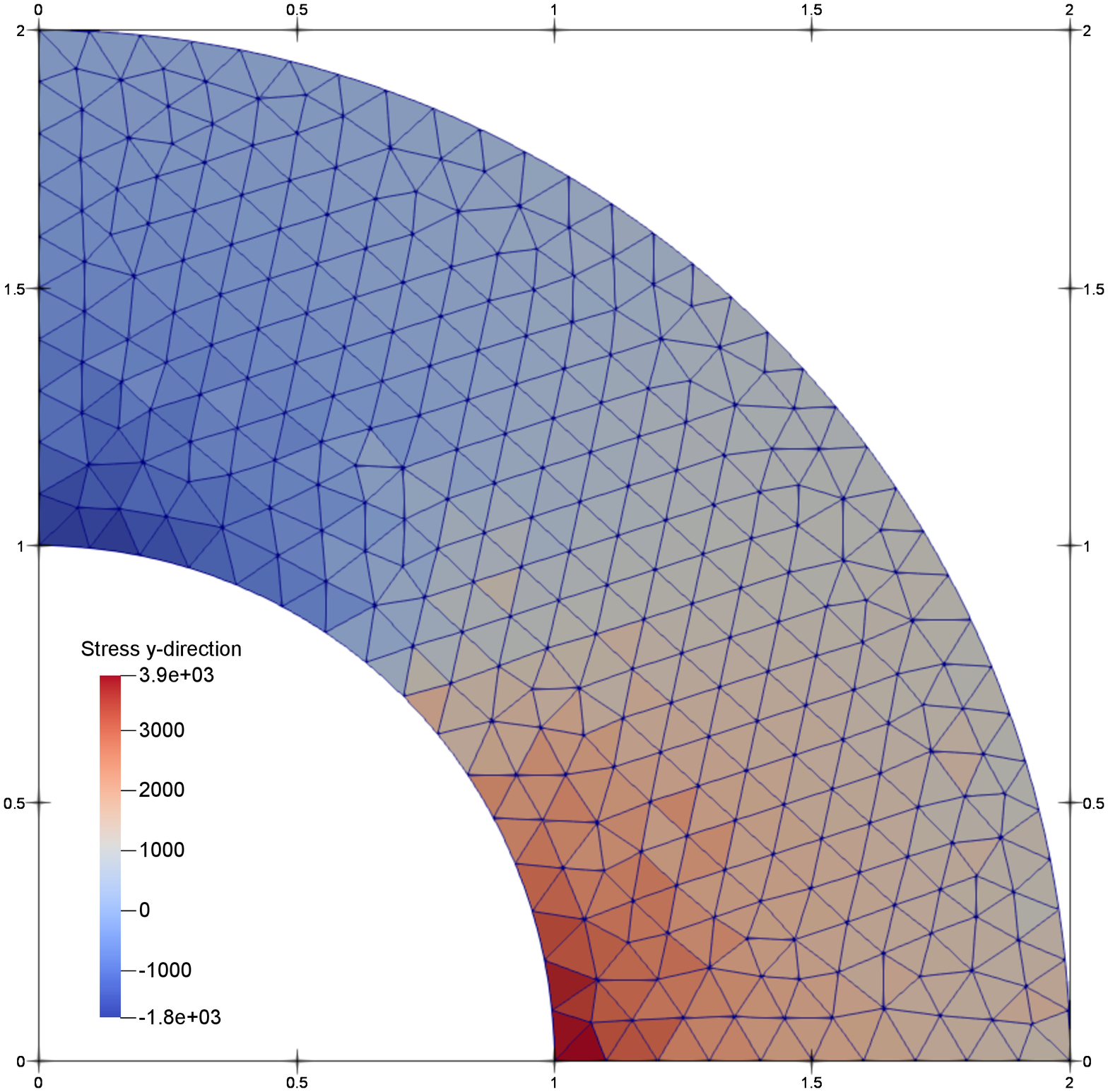} }%
    \label{fig:nl_ddsig_yy}}\\
    \subfloat[]{{\includegraphics[scale=0.26]{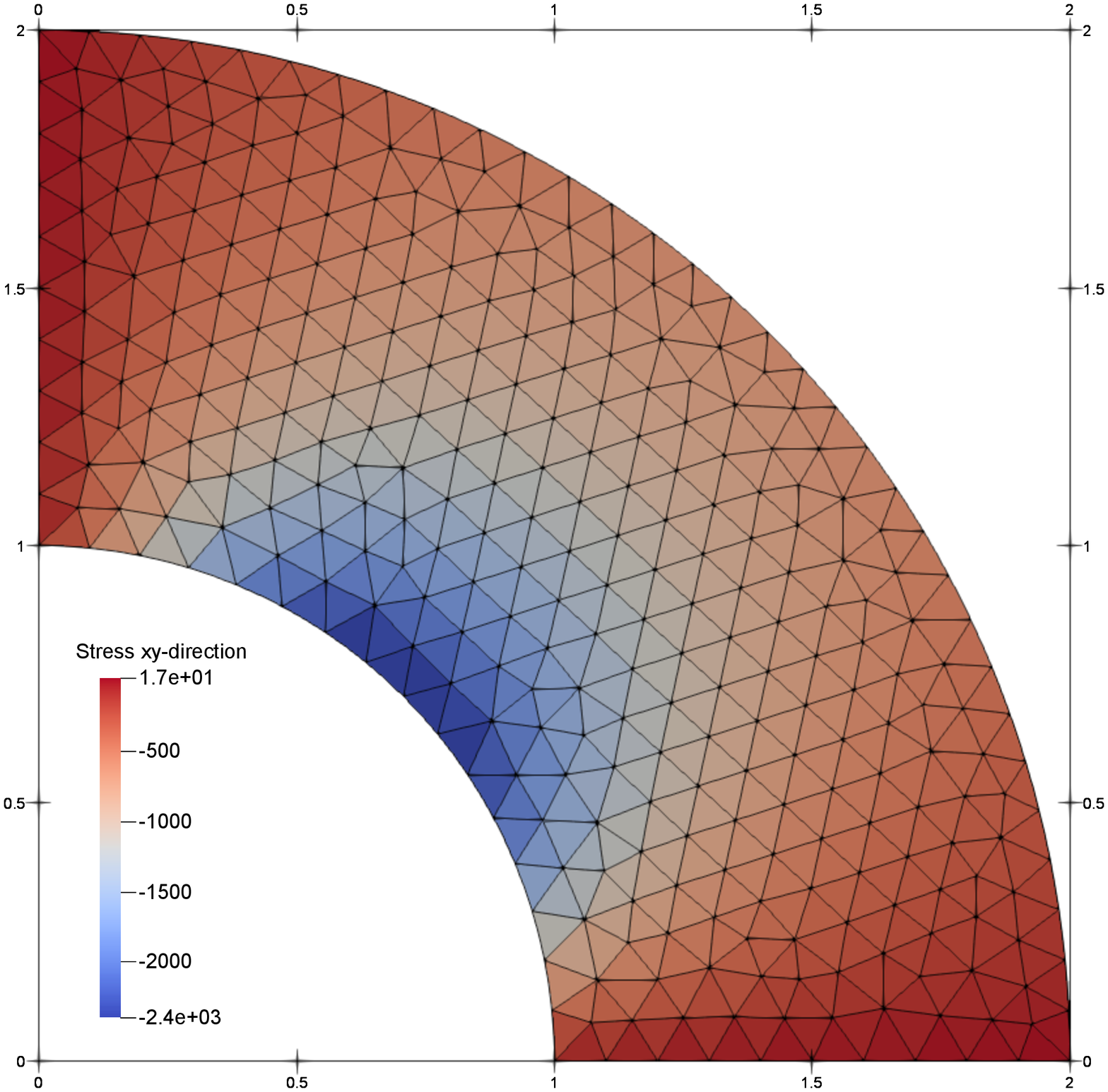} }\label{fig:nl_sig_xy}}%
    \qquad
    \subfloat[]{{\includegraphics[scale=0.26]{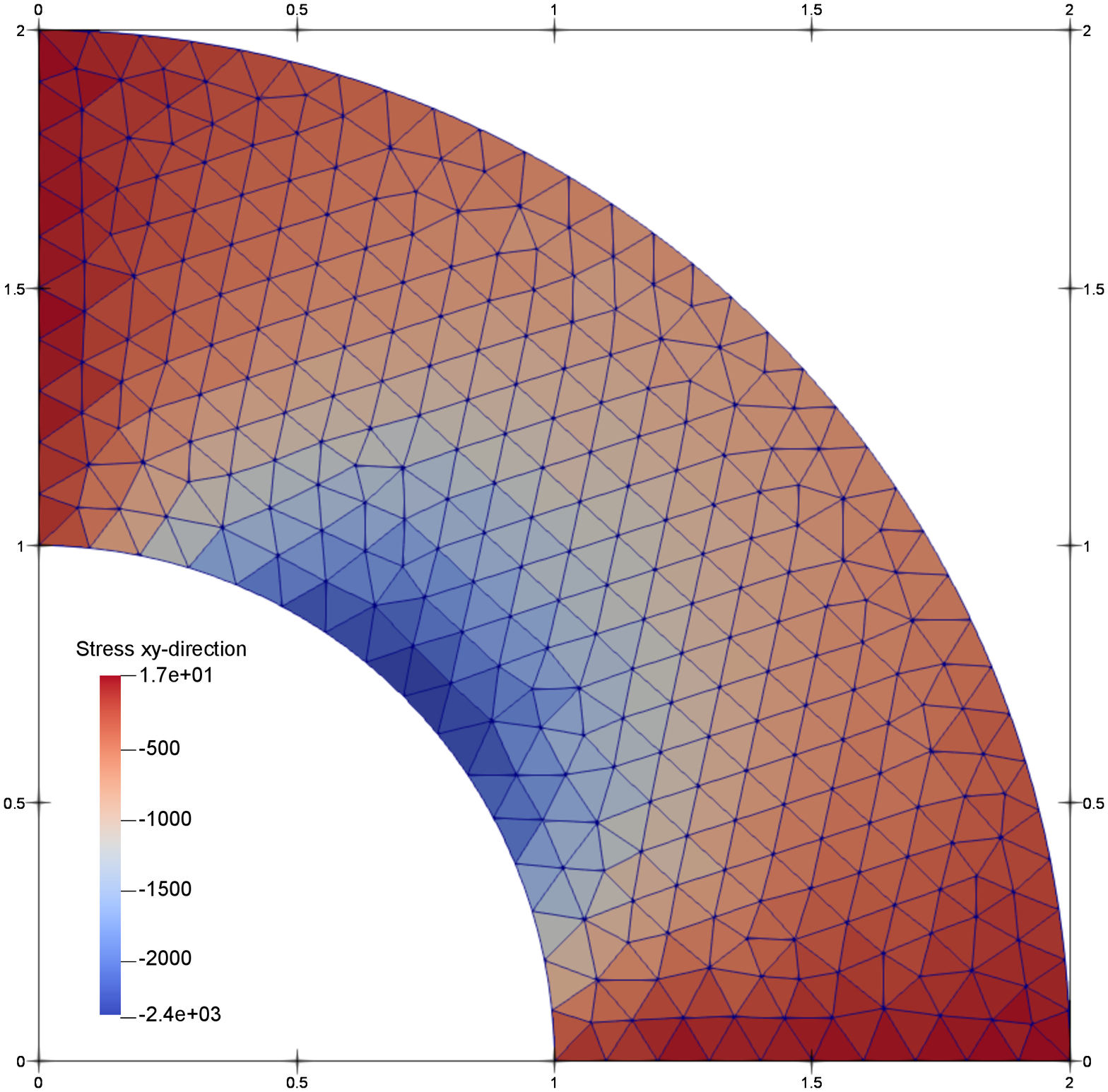} }%
    \label{fig:nl_ddsig_xy}}
    \caption{Comparison of absolute stresses $\fsig [Pa]$ in $x$-direction (a) and (b), in $y$-direction (c) and (d) and in $xy$-direction (e) and (f) between the reference model (left) and the data-driven algorithm (right) based on a normal distribution of size $16^3$. }%
    \label{fig:nl_result}%
\end{figure}

\subsection{Elasto-plastic plate with a circular hole}\label{subsec:exmpl_elasto}
\begin{figure}[H]
\includegraphics[scale=0.28]{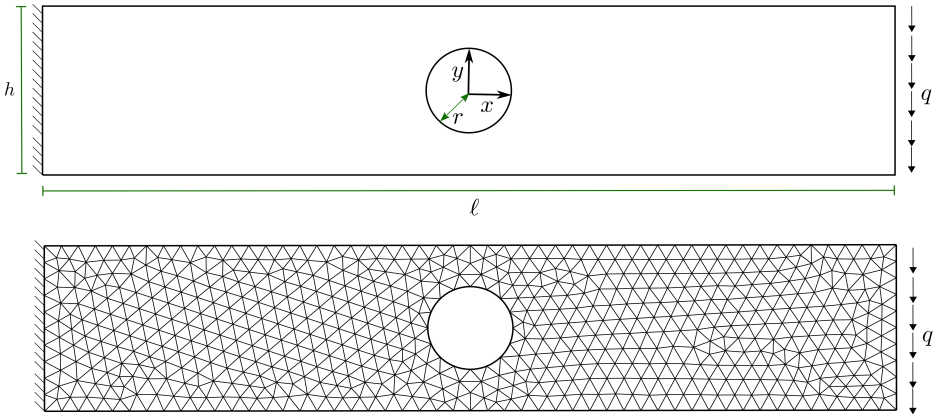}
\caption{Discretization and boundary conditions for a rectangular plate with a circular hole under loading.}
\label{fig:plate_hole}
\end{figure}

\begin{figure}[H]
	\subfloat[a][Reference solution]{\includegraphics[scale=0.4]{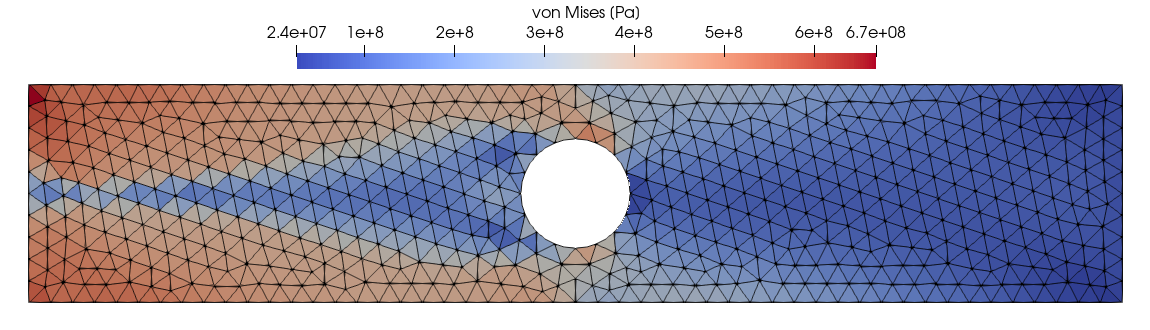} \label{fig:ref_solution}} \vspace{-1cm} \\
	\subfloat[b][Data-driven solution]{\includegraphics[scale=0.4]{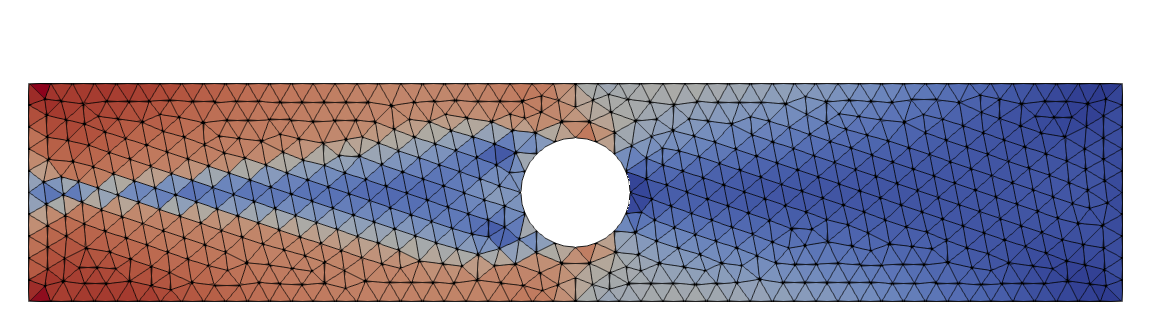} \label{fig:dd_solution}}
	\caption{Von Mises stress distribution at maximum loading at each Gaussian integration point using (a) $J_2$-plasticity model and (b) data-driven algorithm.} \label{fig:vonMisesPlate}
\end{figure}

\begin{figure}[H]
    \subfloat[]{{\includegraphics[scale=0.4]{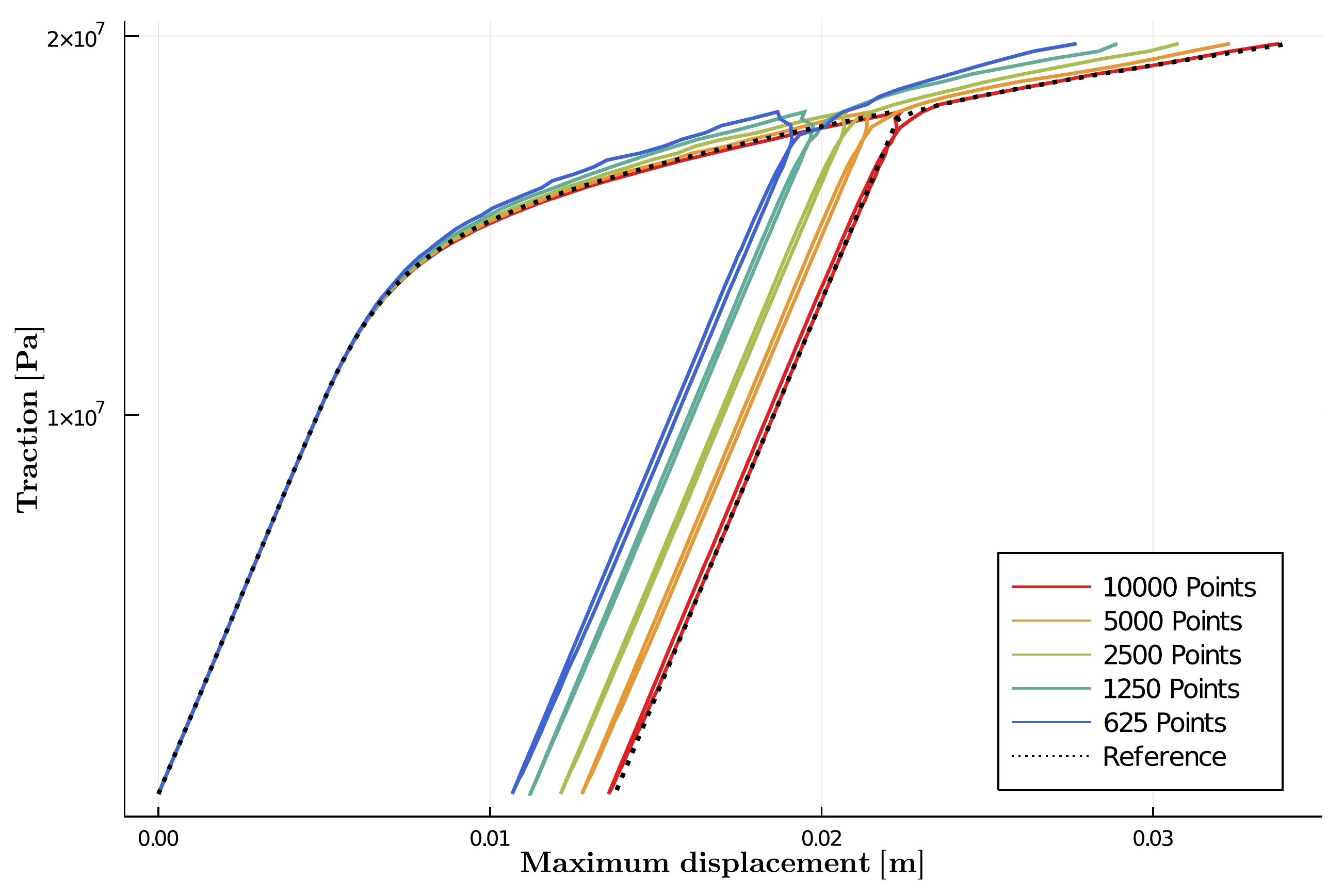} \label{fig:conv_inel_disp}}}%
    \qquad
    \subfloat[]{{\includegraphics[scale=0.45]{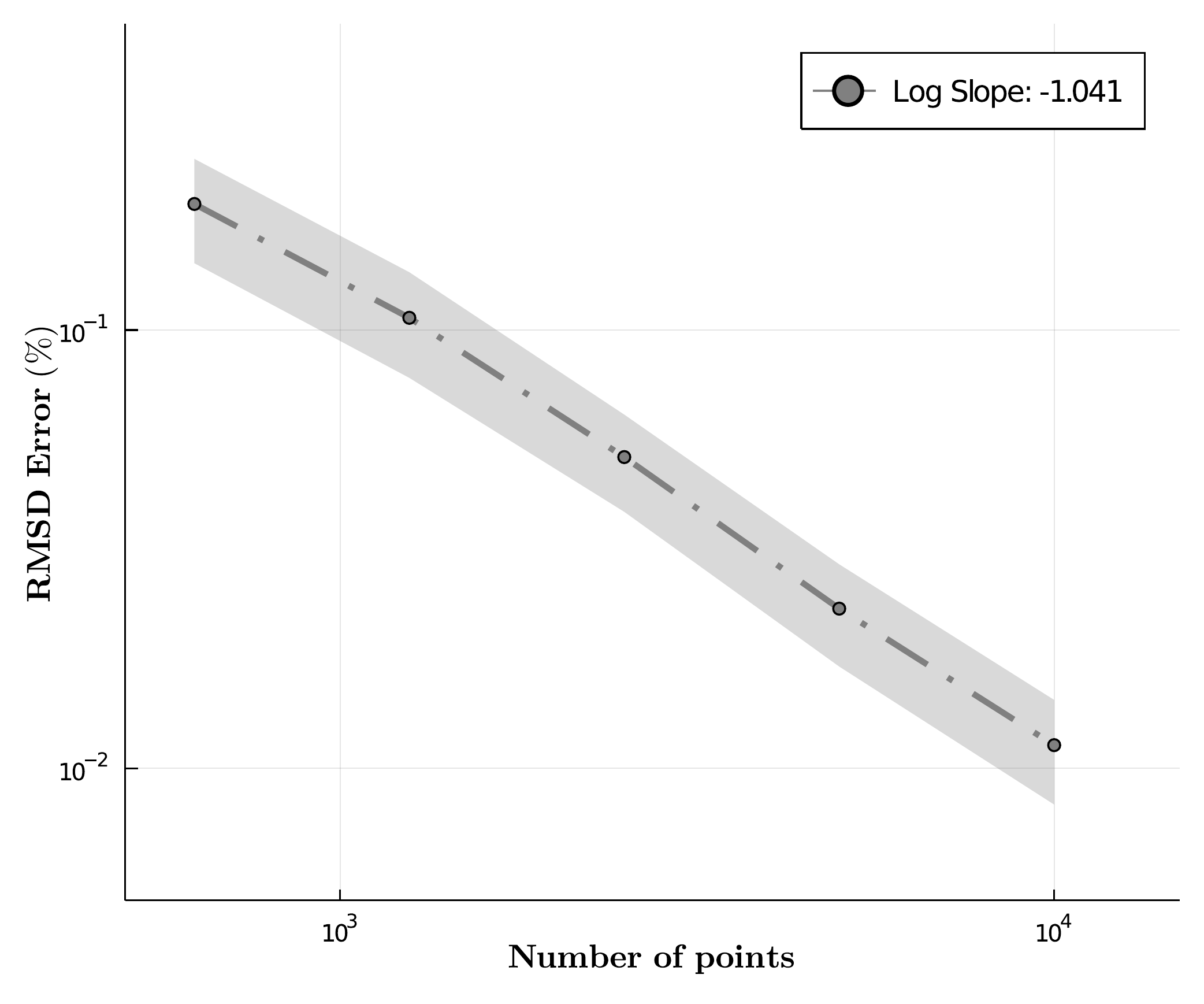} \label{fig:conv_inel_points}}}%
    \caption{Convergence property of the extended data-driven method using transition rules for elasto-plastic material behaviour. (a) Maximum displacement (vertical displacement of lower right vertex versus traction (resultant load of right edge) for different data resolution. (b) RMSD Error for each data resolution. The shaded area shows the deviation of the error arising from different independent virtual tests.}%
    \label{fig:conv_inel}%
\end{figure}

This example illustrates the performance of the data-driven method extended by transition rules by considering an elasto-plastic von Mises material with isotropic hardening for the boundary value problem in Fig.~\ref{fig:plate_hole}. 
The plate with a hole has the dimensions of $\ell = 1\mathrm{m}$, $h = 0.2\,\mathrm{m}$ and $r = 0.05\mathrm{m}$, is clamped at its left edge and subjected to a uniform vertical load $q$ at its right edge. 
The applied load increases from $0$ to $1.8 \cdot 10^7\, \text{Pa}$, decreases to $0$ and then increases again to $2 \cdot 10^7 \, \text{Pa}$, using a constant normalized time step of $\Delta t = 1$.
The material parameters of the reference solid used for the reference solution and data sets are Young's modulus $E = 200\cdot 10^9\, \text{Pa}$, Poisson's ratio $\nu = 0.3$, isotropic hardening modulus $H = E/20$, initial yield stress $\sigma_\mathrm{y0} = 250 \cdot 10^6 \,\text{Pa}$ and elasticity tenssor given by 
\begin{align}\label{eq:elas_matrix2}
\mathbb{C} = (\kappa - \frac{2}{3}G) \fI \otimes \fI + 2G \mathbf{\mathbb{I}},
\end{align}
where $\fI$ is the second-rank identity tensor, $\mathbf{\mathbb{I}}$ is the symmetric part of the fourth-rank identity tensor and $\kappa = \frac{E}{3(1-2\nu)}$ and $G =\frac{E}{2(1+\nu)}$ are the bulk and shear moduli.
The response is computed using a $J_2$-plasticity model based on an iterative return mapping algorithm embedded in a Newton-Raphson global loop restoring equilibrium. 
\begin{figure}
\includegraphics[scale=0.28]{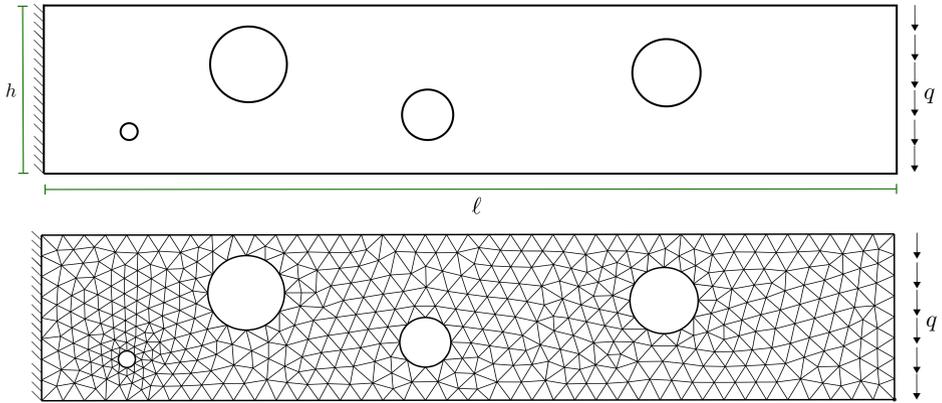}
\caption{Geometry and discretization of an example virtual test of a plate with random holes to generate suitable data sets.}
\label{fig:virtual_geometry}
\end{figure}
Following \cite{eggersmann:2019}, a virtual test employing the geometry depicted in Fig.~\ref{fig:virtual_geometry} is used to generate an accurate coverage of suitable local material states and loading paths of various set sizes. As mentioned before, the corresponding tangents are assumed to be given and therefore calculated analytically using the plasticity model. To ensure uncertainties we endure the total stiffness matrices with some noise. 
Figure~\ref{fig:vonMisesPlate} shows the data-driven solution at the maximum loading magnitude using a data sample containing $10^4$ points.
The convergence of the maximum displacement to the reference displacement based on a $J2$-plasticity model can be seen in Fig.~\ref{fig:conv_inel_disp}.
Moreover, Fig.~\ref{fig:conv_inel_points} confirms a linear convergence rate towards the reference solution by increasing the number of data points.
For better representation, the convergence of the displacement is shown for only one virtual test. However, the convergence of the error is shown for various virtual tests leading to a deviation visualized by the shaded area.
\newpage
\section{Conclusions}\label{sec:conclusion}
We present an approach extending the model-free data-driven computing method of problems in elasticity of Kirchdoerfer and Ortiz \cite{kirchdoerfer:2016} to inelasticity.
The original method uses nearest neighbor clustering and therefore challenges arise dealing with history-dependent data. 
This issue is treated in this work by extending the formulation by including point-wise tangent spaces and classifying the data structure into subsets corresponding to different material behaviour.
Based on the classification, transition rules are defined to map the material point to the classified data subsets, which incorporates with the idea that data points are connected by an underlying structure to each other. 
Additionally, minimizing the distance to local tangent spaces ensures data point connectivity and enables interpolation in regions lacking information of data.
Furthermore, the presented scheme can be easily applied to non-linear elasticity as well, noticing that the resulting system of equations of the minimization problem is reduced, leading to greater efficiency.
A numerical example has been presented to demonstrate the application to data-driven inelasticity and its numerical performance.
Generally, it can be concluded that improvements in accuracy of the presented approach increase for larger data sets and it correlates with the convergence analysis of data-driven elasticity. 
Nevertheless, it should be mentioned that the ensurance of specific quality of the data such as good coverage of material states and loading paths constitutes a critical issue concerning the availability of real experimental data.
Another issue concerns the classification of the data into subsets corresponding to material behaviour. 
This could be done by efficient machine-learning algorithms e.g. spectral or density based clustering.
These generalizations of the data-driven paradigm suggest important directions for future research, especially the usage of machine-learning methods providing further improvement and automation.
%

\bibliography{references}
\bibliographystyle{sn-mathphys}
\end{document}